\documentstyle{europhys}

\newif\ifboo \boofalse

\def\beqa{\begin{eqnarray}}
\def\eeqa{\end{eqnarray}}
\def\beq{\begin{equation}}
\def\eeq{\end{equation}}

\def\eg{{\it e.g. }}

\def\pr{{\it Phys. Rev.}\ }
\def\prl{{\it Phys. Rev. Lett.}\ }
\def\pl{{\it Phys. Lett.}\ }

\def\cqg{{\it Class. Quantum Grav.}\ }

\def\grg{{\it Gen. Relativ. Grav.}\ }

\def\rmp{{\it Rev. Mod. Phys.}\ }

\begin{document}

\euro{??}{??}{??-$\infty$}{1999} \Date{12 January 1999}
\shorttitle{Fermion Helicity Flip Induced by Torsion Field}

\title{Fermion Helicity Flip Induced by Torsion Field }

\author{S. Capozziello\inst{1,3}\footnote{E-Mail:capozziello@vaxsa.csied.unisa.it},
        G.Iovane\inst{1,3}\footnote{E-Mail:geriov@vaxsa.csied.unisa.it},
        G. Lambiase\inst{1,3}\footnote{E-Mail:lambiase@vaxsa.csied.unisa.it} and
        C. Stornaiolo\inst{2,3}\footnote{E-Mail:cosmo.na.infn.it}}
\institute{
     \inst{1} Dipartimento di Scienze Fisiche ``E. R. Caianiello'', Universit\`{a} di Salerno, I-84081 Baronissi, Salerno, Italy.\\
     \inst{2} Dipartimento di Scienze Fisiche, Universit\`{a} di Napoli, Italy.\\
     \inst{3} Istituto Nazionale di Fisica Nucleare, Sezione di Napoli, Italy.}

\rec{}{}

\pacs{ \Pacs{14}{60Pq}{Neutrino mass and mixing}
\Pacs{95}{30Sf}{Relativity and gravitation}
\Pacs{04}{50+h}{Alternative theories of gravity}}

\maketitle

\begin{abstract}
We show that in theories of gravitation with torsion the helicity
of fermion particles is not conserved and we calculate the
probability of spin flip, which is related to the anti-symmetric
part of affine connection. Some cosmological consequences are
discussed.
\end{abstract}

\section{Introduction}
Attempts to conciliate General Relativity with Quantum Theory
yielded to propose theories of gravitation including torsion
fields, so that the natural arena is the space--time $U_4$ that is
a generalization of Riemann manifold $V_4$.

The advantage to pass from $V_4$ to $U_4$ is due to the fact that
the spin of a particle turns out to be related to the torsion just
as its mass is responsable of the curvature. From this point of
view, such a generalization tries to include the spin fields of
matter into the same geometrical scheme of General Relativity.

One of the attempts in this direction is the
Einstein--Cartan--Sciama--Kibble (ECSK) theory \cite{hehl}.
However the torsion seems to play an important role in any
fundamental theory. For instance: a torsion field appears in
(super)string theory if we consider string fundamental modes; we
need, at least, a scalar mode and two tensor modes: a symmetric
and antisymmetric one. The latter, in the low energy limit for
string effective action, gives the effects of a torsion field
\cite{GSW}; any attempts of unification between gravity and
electromagnetism require the inclusion  torsion in four and in
higher--dimensional theories as Kaluza--Klein ones \cite{GER};
theories of gravity formulated in terms of twistors need
 the inclusion of torsion \cite{HOW};
in the supergravity theory torsion, curvature and matter fields
are treated under the same standard \cite{LOS}; in cosmology
torsion could have had a relevant role into dynamics of the early
universe because it gives a repulsive contribution to the
energy--momentum tensor so that cosmological models become
singularity--free \cite{SIV}, and if the universe undergoes one or
several phase transitions, torsion could give rise to topological
defects (\eg torsion walls \cite{FIG}) which today can result as
intrinsic angular momenta for cosmic structures as galaxies.

Some macroscopic observable effects of torsion in the framework of
cosmology has been studied in Ref. \cite{ZZI} where it is shown
that the presence of torsion into effective energy--momentum
tensor alters the spectrum of cosmological perturbations giving
characteristic lengths for large scale structures. As a final
remark, we have to note that  spacetime torsion, being related to
the intrinsic spin degrees of freedom of matter \cite{hehl},
cannot be transformed away, so that we have to expect its remnants
at any epoch of cosmological evolution.

All these arguments do not allow to neglect torsion in any
comprehensive theory of gravity which takes into account
non-gravitational counterpart of fundamental interactions.

The purpose of this paper is to show that, in presence of torsion,
the helicity of fermion particles is not conserved. This effects
could be important for testing some astrophysical consequences of
torsion \cite{HAM} because of smallness of coupling constant with
respect to the other fundamental interactions.

Our starting point is to consider the Dirac equation in the
space--time $U_4$. Due to the torsion, it acquires an additional
coupling term of the form
$(1/4)S_{\alpha\beta\sigma}\gamma^{\alpha}\gamma^{\beta}\gamma^{\sigma}$,
where $S_{\alpha\beta\sigma}$ is related to the antisymmetric part
of the affine connection, $\Gamma^{\sigma}_{[\alpha,
\beta]}=S_{\alpha\beta}^{\sigma}$. This term is, as we will see,
responsable of the spin flipping of fermions.

It is worthwhile to note that helicity flips are induced also by
gravitational fields, as consequence of coupling between spin and
curvature \cite{PAP}.

The paper is organized as follows. In Section 2 we will shortly
review the basic concepts leading to the Dirac equation in
presence of torsion fied. In Section 3 we show that the helicity
operator of a fermionc particle is not conserved. The probability
that the flip helicity occurs is calculated in Section 4.
Conclusions are discussed in Section 5.

\section{The Dirac Hamiltonian}
The Dirac equation in curved space--time is written in terms of
the vierbeins formalism \cite{BIR}. One introduces the vierbein
fileds $e_{\mu}^a(x)$ where the Latin indices refer to the locally
inertial frame and Greek indices to a generic non--inertial frame.
The non--holonomic index $a$ labels the vierbein, while the
holonomic index $\mu$ labels the components of a given vierbein.
The connection in non--holonomic coordinates is given by
\cite{HAM}
\begin{equation}\label{eq2.1}
\Gamma_{abc}=-\Omega_{abc}+\Omega_{bca}-\Omega_{cab}+S_{abc}\,{,}
\end{equation}
where $\Omega_{\alpha\beta}^c\equiv e^c_{[\alpha,\beta]}$,
$\Omega^c_{ab}= e_a^{\alpha}\,e_b^{\beta}\,c_{\sigma}^c\,
\Omega_{\alpha\beta}^{\sigma}$, and $S_{abc}$ is the
anti-symmetric part of the affine connection. The covariant
derivative is defined as
\begin{equation}\label{eq2.2}
D_{\mu}\equiv \partial_{\mu}-\frac{1}{4}\Gamma_{\mu
ab}\gamma^a\gamma^b\,{,}
\end{equation}
and the Dirac equation is given by
\begin{equation}\label{eq2.3}
\gamma^a D_a\psi+i\frac{mc}{\hbar}\psi=0\,{.}
\end{equation}
In the spirit to study only the effects due to the torsion, we
will neglect  gravitational effects to the spin flip (they have
been analyzed in details in Ref. \cite{PAP}). It means to neglect
the $\Omega_{abc}$ terms in eq. (\ref{eq2.1}) so that the Dirac
equations assumes the form
\begin{equation}\label{eq2.4}
\gamma^{\alpha}\psi_{,\alpha}+i\frac{mc}{\hbar}\psi=\frac{1}{4}
S_{\alpha\beta\sigma}\gamma^{\alpha}\gamma^{\beta}\gamma^{\sigma}\psi
\end{equation}
>From it one derives the Hamiltonian
\begin{equation}\label{eq2.5}
H=c\vec{\alpha}\cdot\vec{p}+mc^2\beta+\frac{i}{4}S_{\alpha\beta\sigma}
\gamma^0\gamma^{\alpha}\gamma^{\beta}\gamma^{\sigma}=H_0+H^{\prime}\,{,}
\end{equation}
where $H^{\prime}$ is a perturbation of the unperturbed
Hamiltonian $H_0=c\vec{\alpha}\cdot\vec{p}+mc^2\beta$.

\section{Helicity flip of fermions}
In this section we will prove that the helicity of a fermion is
not conserved in a space $U_4$. This follows by calculating the
time variation of the helicity operator in the Heisenberg
representation and showing that it does not vanish.

The helicity operator is defined as \cite{ITZ}
\begin{equation}\label{eq3.1}
h=\vec{\Sigma}\cdot\vec{\hat{p}}\,{,}
\end{equation}
where the spin matrix $\vec{\Sigma}$ and the versor
$\vec{\hat{p}}$ are
\begin{equation}\label{eq3.2}
\Sigma^i= \left( \begin{array}{cc}
              \sigma^i & 0 \\
              0 & \sigma^i \end{array} \right)  \,{,}
\quad \hat{p}^i=\frac{p^i}{|\vec{p}|}\,{.}
\end{equation}
$\sigma^i, i=1,2,3$ are the Pauli matrices and $p^{\mu}=(p^0,
\vec{p})$ is the momentum. In the Heisenberg representation the
dynamical evolution of the helicity operator is given by
\begin{equation}\label{eq3.3}
i\hbar\dot{h}=[h,H]\,{,}
\end{equation}
where $H$ is the Hamiltonian of the system under consideration.
For the Hamiltonian (\ref{eq2.5}) one gets
\begin{equation}\label{eq3.4}
i\dot{h}=\frac{cp^k}{4|\vec{p}|}\varepsilon_{ijk}S_{\alpha\beta\sigma}
\gamma^0\left[ g^{i\sigma}\gamma^{\alpha}\gamma^{\beta}\gamma^j+
2g^{i\alpha}g^{j\beta}\gamma^{\sigma}\right] \,{.}
\end{equation}
Eq. (\ref{eq3.4}) implies that $\dot{h}\neq 0$ so that the
helicity of fermion particle is not conserved.

\section{Probability of spin flipping}
In this Section, we will calculate the probability of the helicity
flip induced by the torsion term in Eq. (\ref{eq2.5}). We consider
the totally anti-symmetric dual or a null vector, $S^{\sigma}=
(|\vec{S}|, \vec{S})$. We also used the approximation
$g_{\mu\nu}\sim \eta_{\mu\nu}$. Then the Hamiltonian (\ref{eq2.5})
can be recast in the form
\begin{equation}\label{eq4.1a}
H^{\prime}=-\frac{3\hbar c|\vec{S}|}{2}\gamma^5+i\frac{3}{2}
\vec{S}\cdot\left( \begin{array}{cc}
   \vec{\sigma} & 0 \\
       0  & \vec{\sigma} \end{array} \right) \,{.}
\end{equation}
where $\gamma^5$ is defined as \cite{ITZ}
\begin{equation}\label{gamma5}
\gamma^5=i\gamma^0\gamma^1\gamma^2\gamma^3= \left(
\begin{array}{cc}
              0 & 1 \\
              1 & 0 \end{array} \right)  \,{.}
\end{equation}

The state of a fermion particle is described by spinor $$\psi
(x)=\left( \begin{array}{c} \psi_R \\ \psi_L \end{array}
\right)\,{,}$$ so that it can be rewritten as a superposition of
states $|\psi_R>$ and $|\psi_L>$. For istance, at $t=0$ one has
\begin{equation}\label{eq4.1}
|\psi(0)>=a_0|\psi_R>+b_0|\psi_L>\,{,}
\end{equation}
where $a_0, b_0$ are constants, $|\psi_R>$ and $|\psi_L>$ are
eigenkets of energy, i.e. $H_0|\psi_{R/L}>=E|\psi_{R/L}>$. We
choose the independent kets $$ |\psi_R>\equiv \left(
\begin{array}{c} 1 \\ 0 \end{array} \right)\,{,} \quad
|\psi_L>\equiv \left( \begin{array}{c} 0 \\ 1 \end{array}
\right)\,{.} $$ The time evolution of the state (\ref{eq4.1})
\begin{equation}\label{eq4.2}
|\psi(t)>=a(t)|\psi_R>+b(t)|\psi_L>\,{,}
\end{equation}
is given by recasting Dirac's equation as a Schr\"odinger like one
\begin{equation}\label{eq4.3}
i\hbar\frac{\partial}{\partial t}|\psi
(t)>=(H_0+H^{\prime})|\psi(t)>\,{.}
\end{equation}
Inserting Eq. (\ref{eq4.2}) into Eq. (\ref{eq4.3}), at first order
of perturbative calculation, one gets
\begin{equation}\label{eq4.4}
i\hbar\frac{\partial}{\partial t} \left( \begin{array}{c} a \\ b
\end{array} \right) =M \left( \begin{array}{c} a \\ b \end{array}
\right)\,{.}
\end{equation}
where $M$ is the matrix
\begin{equation}\label{eq4.5}
M=\left( \begin{array}{cc}
   <\psi_R|H_0+H^{\prime}|\psi_R> & <\psi_R|H^{\prime}|\psi_L> \\
<\psi_L|H^{\prime}|\psi_R> & <\psi_L|H_0+H^{\prime}|\psi_L>
\end{array} \right) \,{.}
\end{equation}
Explicit calculation of the matrix elements yields
\begin{equation}\label{eq4.6}
M=\left( \begin{array}{cc}
         E +i3\hbar c|\vec{S}|^2 & \hbar c|\vec{S}| \\
   \hbar c|\vec{S}| & E+i3\hbar c|\vec{S}|^2 \end{array} \right) \,{.}
\end{equation}
By diagonalizing the matrix (\ref{eq4.6}), one derives the
eigenvalues
\begin{equation}\label{eq4.7}
\lambda_{\pm}=E+i3|\vec{S}|^2\pm \frac{3\hbar c}{2}|\vec{S}|\,{,}
\end{equation}
and the corresponding normalized eigenkets
\begin{equation}\label{eq4.8}
|\lambda_{\pm}>=\frac{1}{\sqrt{2}}[|\psi_R>\pm\eta_{\pm}|\psi_L>]\,{.}
\end{equation}
In Eq. (\ref{eq4.8}), $\eta_{\pm}$ are the phase factors that we
choose to be equal to one. It implies that at $t=0$, $|\psi
(0)>=|\psi_R>$, i.e. $a_0=1, b_0=0$ in the Eq. (\ref{eq4.1}).
Then, the evolution of the state $|\psi (t)>$ can be written as
\begin{eqnarray}\label{eq4.9}
|\psi (t)> & = & \frac{1}{\sqrt{2}}[e^{-i\lambda_+
t/\hbar}|\lambda_+> +e^{-i\lambda_- t/\hbar}|\lambda_- >]= \\
       & = & e^{-i(E/\hbar+3c|\vec{S}|/4) t}\,
e^{- 3c|\vec{S}|^2 t}\, [\cos\frac{c\vert S\vert}{2}t\, |\psi_R> +
\sin\frac{c\vert S\vert }{2}t\,|\psi_L>]\,{.} \nonumber
\end{eqnarray}
Eq. (\ref{eq4.9}) describes the state of a fermion at time $t$ if
it starts as $|\psi_R>$. The probability to find it in state
$|\psi_R>$ at time $t$ is $P_R(t)\sim\cos^2(3c|\vec{S}|/4)t$,
while the probability that the spin flip occurs is
$P_L(t)\sim\sin^2(3c|\vec{S}/4)t$.

The frequency of spin flipping is $\omega=3c|\vec{S}|/4$, from
which follows the characteristic length $L=8\pi/3|\vec{S}|$.

Due to the dissipative term, the state decrease exponentially.
This fact has important consequences in the very early universe.

\section{Conclusions}
In this paper, we calculate the probability that a background
torsion source induces a spin flip on fermion particles moving in
it. The torsion field is described by a null vector. We are
dealing with high energy fermion particles, so that helicity can
be identified with spin; this method can work both for fermion
massive and massless particles.

This phenomenology can occur in a regime where the effects of
torsion become of the same order of magnitude or bigger than those
due to energy momentum tensor at extremely high densities and at
sufficiently high polarization of fermion particles. Such a
scenario could realize at early cosmological epoch where particle
density becomes similar to the critical cosmological density; for
example, this happens if electrons are taken into account and
$kT\simeq 10^{11}$GeV \cite{BAU}. It means that, at this epoch,
the probability $P_L(t)$ has to be different from zero. In this
sense, torsion and spin density can assume  relevant roles in the
today observed astrophysical structures, resulting, for example,
as intrinsic macroscopic angular momenta \cite{FIG}.

\end{document}